\documentclass{moriond}

\def\be{\begin{equation}}
\def\ee{\end{equation}}
\def\bea{\begin{eqnarray}}
\def\eea{\end{eqnarray}}

\newcommand{\Photo}{\includegraphics[height=35mm]{mypicture}}

\begin{document}
\vspace*{4cm}
\title{Angular ordering effects in TMD parton distribution functions and Drell-Yan $q_{\bot}$ spectra\\
\footnotesize{Contribution to the 2019 QCD session of the 54th Rencontres de Moriond}}


        



\author
{Aleksandra Lelek\\
\normalsize{ University of Antwerp (UAntwerp)}\\
}


\maketitle\abstracts{
We present new results of our studies of soft-gluon angular ordering effects on the evolution of both collinear and transverse momentum dependent (TMD) parton distribution functions, and discuss their implications for precision predictions of Drell-Yan transverse momentum spectra at the LHC. Our method is based on the parton branching (PB) approach. We compare this with the implementation of angular ordering in the Kimber-Martin-Ryskin-Watt (KMRW) approach and with the Collins-Soper-Sterman (CSS) approach.  We illustrate numerically the effects of different ordering scenarios ($p_{\bot}$, angular ordering), including definitions of the soft-gluon resolution scale and scale in the running coupling, on the theoretical accuracy of predictions in the low transverse momentum region of Drell-Yan spectra measured at the LHC.}

\section{Motivation}


One of the uncertainty sources in  obtaining QCD predictions for collider measurements comes from the assumption that partons are collinear with the hadron they build. The collinear factorization theorem \cite{Collins:1989gx} successfully predicts a large number of processes. However, it was realised a long time ago that for certain observables also the parton's transverse momentum needs to be taken into account in order to obtain sufficient precision. This is accomplished via transverse momentum dependent (TMD) factorization theorems, like  high energy $k_{\bot}$- factorization \cite{Catani:1990xk,Catani:1990eg} or Collins-Soper-Sterman \cite{Collins:1984kg} formalism.
An overview of the field is given in \cite{Angeles-Martinez:2015sea}.

In the following sections we present new results from the  Parton Branching (PB) method \cite{Hautmann:2017xtx,Hautmann:2017fcj,Martinez:2018jxt,Lelek:2018vtr,Hautmann:2019biw} to obtain TMD parton distribution functions (PDFs), referred to as TMDs. We concentrate especially on the proper treatment of soft gluons emissions via the angular ordering condition and on the comparison of PB with other existing approaches \cite{Hautmann:2019biw}, as Marchesini's and Webber's  prescription \cite{Marchesini:1987cf}, Kimber-Martin-Ryskin-Watt (KMRW) approach \cite{Kimber:1999xc,Martin:2009ii} and Collins-Soper-Sterman \cite{Collins:1984kg} formalism.

\section{TMDs from PB method}
\subsection{TMD evolution equation}
The PB TMDs are obtained by constructing and solving using MC techniques an evolution equation which takes into account not only the collinear evolution but also the transverse momentum at each branching. The proposed equation has the following form \cite{Hautmann:2017fcj}
\begin{eqnarray}
&& \widetilde{A}_a\left( x, k_{\bot}, \mu^2\right) = \Delta_a\left(\mu^2, \mu_0^{2}\right)\widetilde{A}_a\left( x, k_{\bot}, \mu_0^2\right)+ 
 \sum_b\int \frac{\textrm{d}^2\mu_{\bot}^{\prime}}{\pi \mu^{\prime 2}}\Theta\left(\mu^{2}-\mu^{\prime 2}\right)\Theta\left(\mu^{\prime 2}-\mu_0^{ 2}\right)\times \nonumber \\ &\times& \Delta_a\left(\mu^2, \mu^{\prime 2}\right)
  \int_x^{z_M}\textrm{d}z P_{ab}^{R}\left(z,\alpha_s(a(z)^2\mu^{\prime 2})\right)\widetilde{A}_b\left( \frac{x}{z}, k_{\bot} + a(z)\mu_{\bot}, \mu^{\prime 2}\right) 
  \label{eq:tmdevol}
\end{eqnarray} 
where $\widetilde{A}_a\left( x, k_{\bot}, \mu^2\right)= x A_a\left( x, k_{\bot}, \mu^2\right)$ is the momentum weighted TMD for a  parton species and flavour $a$, carrying the fraction $x$ of the proton's momentum and having the transverse momentum $k_{\bot}$  \footnote{For a given $4$-vector  $k=(k^0, k^1, k^2, k^3)=(E_{k}, k_{\bot}, k^3)$, where $k_{\bot}=(k^1, k^2)$. Analogously $\mu_{\bot}^{\prime}=(\mu^{1 \prime}, \mu^{2 \prime} )$.} at the evolution scale $\mu$, $P_{ab}^R$ is the real-emission part of the splitting function for a parton $b$ splitting into a parton $a$ which propagates towards the hard scattering, $z = x_a/x_b$ is the splitting variable,  $|\mu_{\bot}|\equiv \mu^{\prime } $ is the scale at which the branching happens, $\mu_0$ is the initial evolution scale.  $\Delta_a(\mu^2, \mu_0^2) = \exp\left[-\int_{\mu_0^2}^{\mu^2}\frac{\textrm{d}\mu^{\prime 2}}{\mu^{\prime 2}}\sum_b \int_0^{z_M}\textrm{d}z z P_{ba}^{R}\left(z,\alpha_s\left(a(z)^2\mu^{\prime 2}\right)\right) \right]$ is the Sudakov form factor. The function $a(z)$ expresses the relation between the scale of the branching and the transverse momentum of the emitted and propagating parton. For $p_{\bot}$-ordering condition, $a(z)=1$, i.e. the scale of the branching is associated with the transverse momentum of the emitted parton  $q_{\bot}^2 = \mu_{\bot}^{\prime 2}$. For angular ordering condition, $a(z) = 1-z$, i.e. the scale of the branching is associated with energy times the angle of the emitted parton with respect to the beam direction $q_{\bot}^2 = (1-z)^2\mu_{\bot}^{\prime 2}$. The soft gluon resolution scale parameter $z_M$ is fixed to a value very close to $1$ for  $p_{\bot}$- ordering or it is defined as  $z_M = 1-\frac{q_0}{\mu^{\prime}}$ for angular ordering where $q_0$ is the minimum transverse momentum of the emitted parton with which it can be resolved. The PB method allows one to select the definition of $z_M$, $\alpha_s$ and the way the transverse momentum is related to the branching scale independently \footnote{E.g. one can select angular ordered way of relating $q_{\bot}$ and $\mu^{\prime}$ but keeping $z_M$ fixed and $\alpha_s(\mu^{\prime 2})$ as in $p_{\bot}$-ordering to study the effect of each piece of the ordering definition. }. \\
In the PB method the transverse momentum of the propagating parton is a sum of the transverse momentum of all the emitted partons 
$
k_{\bot} = -\sum_i q_{\bot, i}\;.
$
After integrating eq.~(\ref{eq:tmdevol}) over the transverse momentum $k_{\bot}$ one obtains collinear PDF. In the limit of $z_M \rightarrow 1$ and with $\alpha_s(\mu^{\prime 2})$  the Dokshitzer-Gribov-Lipatov-Altarelli-Parisi (DGLAP) evolution equation \cite{Gribov:1972ri,Lipatov:1974qm,Altarelli:1977zs,Dokshitzer:1977sg} is reproduced. 
\subsection{Highlights}
\begin{figure}
\begin{minipage}{0.3\linewidth}
\centerline{\includegraphics[width=0.9\linewidth]{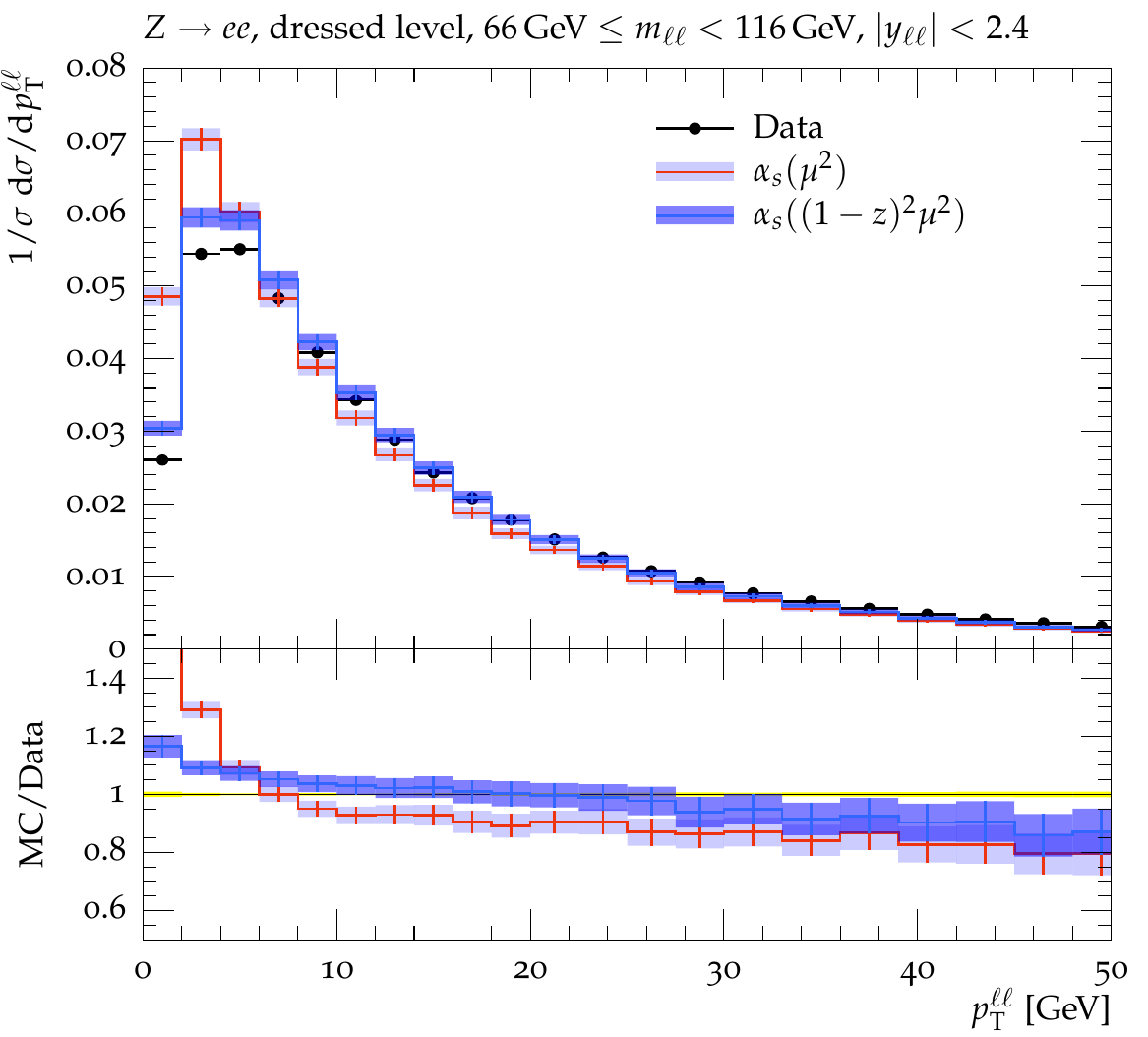}}
\end{minipage}
\begin{minipage}{0.3\linewidth}
\centerline{\includegraphics[width=0.99\linewidth]{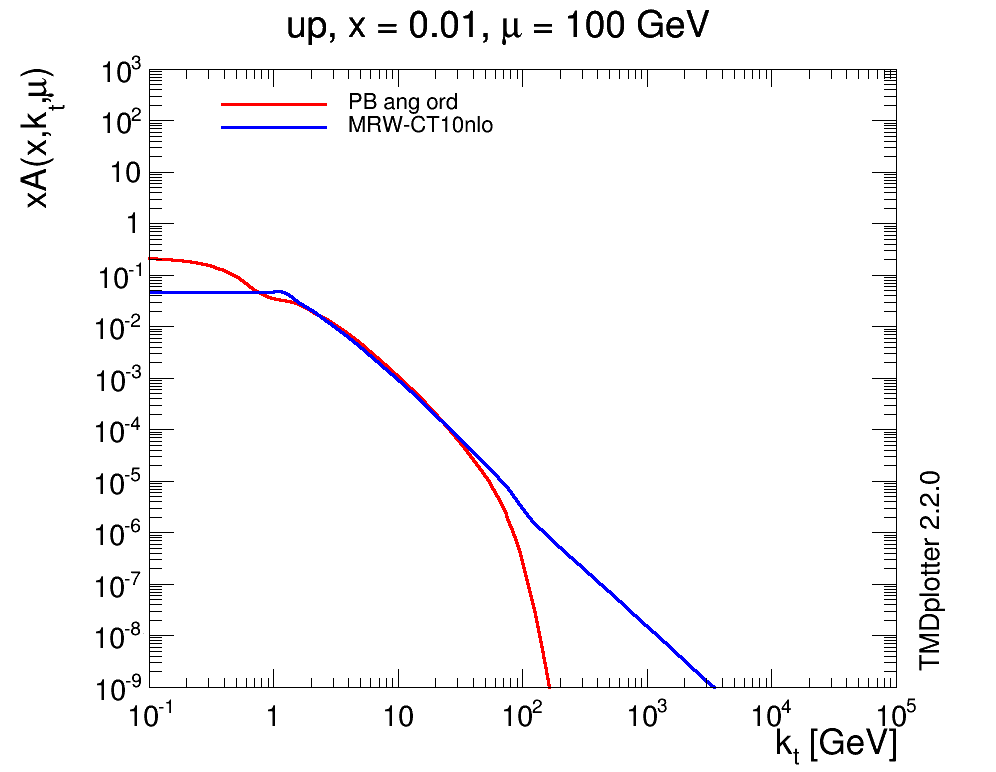}}
\end{minipage}
\hfill
\begin{minipage}{0.3\linewidth}
\centerline{\includegraphics[width=0.99\linewidth]{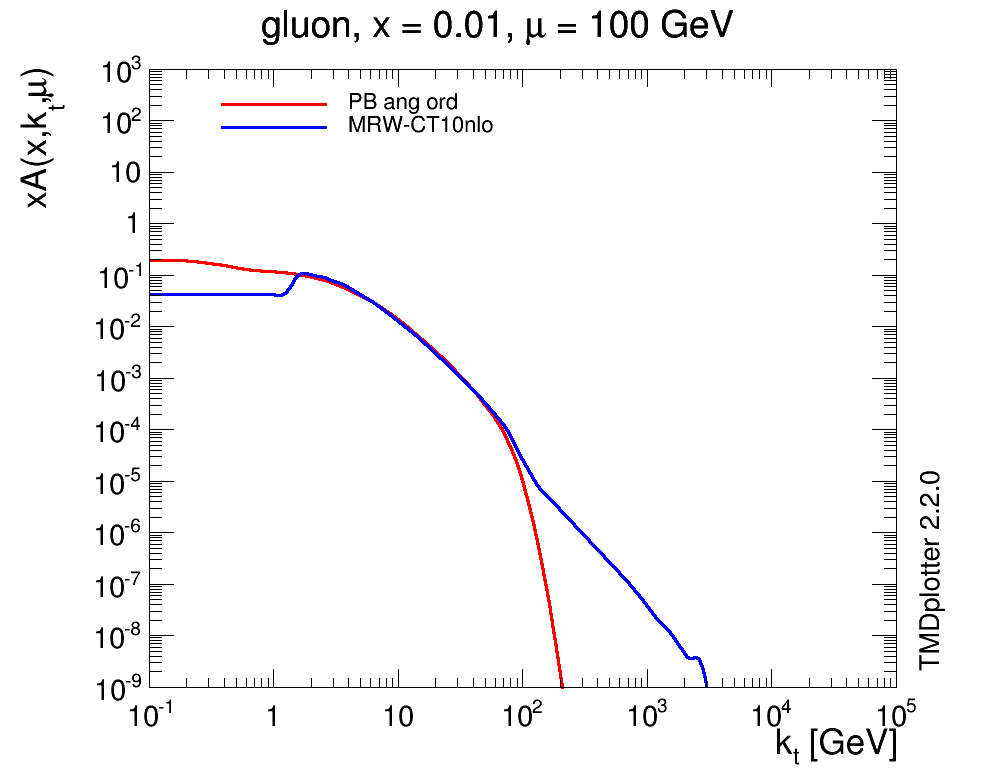}}
\end{minipage}
\hfill
\caption[]{ Prediction for Z boson $p_{\bot}$ spectrum obtained with PB TMDs from the fit compared to 8 TeV ATLAS measurement (left). Comparison of KMRW and PB TMD (middle and right).   }
\label{fig:KMRWvsPB}
\end{figure}

The key observation was that if one relates the transverse momentum and the scale of the branching according to the angular ordering condition, stable, $z_M$ independent (as long as $z_M\approx 1$ ) TMD is obtained whereas with $p_{\bot}$ ordering it is not possible \cite{Hautmann:2017xtx,Hautmann:2017fcj}. Based on this result, fits of TMDs to  precision measurements of deep inelastic scattering (DIS) cross sections at HERA were performed using \texttt{xFitter} \cite{Alekhin:2014irh} for angular ordering ($q_{\bot}^2 = (1-z)^2\mu_{\bot}^{\prime 2}$) in two scenarios: with $\alpha_s(\mu^{\prime 2})$ and $\alpha_s(q_{\bot}^{ 2})$   \cite{Martinez:2018jxt}. A very good description of the $Z$ boson $p_{\bot}$ spectrum measured by ATLAS experiment at 8 TeV \cite{Aad:2015auj} was obtained with $\alpha_s(q_{\bot}^{ 2})$  \cite{Martinez:2018jxt,Martinez:2019mwt}which is shown in the left panel of fig.~\ref{fig:KMRWvsPB} \cite{Martinez:2018jxt}.

\section{PB and other approaches}

\subsection{Marchesini and Webber}
Eq.~(\ref{eq:tmdevol}) with angular ordering condition, once  integrated over transverse momentum, gives the following evolution formula for the collinear distribution
\begin{eqnarray}\label{eq:PBangular}
 && \widetilde{f}_{a}(x,\mu ^{2}) = \widetilde{f}_{a}(x,\mu _{0} ^{2})\Delta_a(\mu ^{2}, \mu_0 ^{2})  \nonumber \\ &&+  \int_{\mu_{0} ^{2}}^{ \mu ^{2}}  \frac{d \mu ^{\prime 2}}{\mu ^{\prime 2}}\Delta_a(\mu ^{2},\mu ^{\prime 2}) \sum _{b} \int _{x}^{1-\frac{q_0}{\mu^{\prime}}} dz P^{R}_{ab}\left(\alpha_s\left((1-z)^2\mu^{\prime 2}\right), z\right)\widetilde{f}_{b}\left(\frac{x}{z},\mu ^{\prime 2}\right)\;.
\end{eqnarray} 
This coincides with the evolution formula of Marchesini and Webber \cite{Marchesini:1987cf} \footnote{Note, that Marchesini and Webber studied the coherent branching with LO splitting functions and $\alpha_s$ whereas we are using them at NLO.}.

\subsection{Kimber-Martin-Ryskin-Watt}
In this section we compare the PB formula with the KMRW approach \cite{Kimber:1999xc,Martin:2009ii}. To this end, we rewrite the PB formula for angular ordering eq.~(\ref{eq:PBangular}) in terms of integral over the transverse momentum $q_{\bot}^2$ instead of the branching scale $\mu^{\prime 2}$ \footnote{We neglect the difference between $\mu_0$ and $q_0$.}
\begin{eqnarray}\label{PBangular_2term_3}
 && \widetilde{f}_{a}(x,\mu ^{2}) = \widetilde{f}_{a}(x,\mu _{0} ^{2})\Delta_a(\mu ^{2}, \mu_0^2)  \nonumber \\&&+\int_{q_{0} ^{2}}^{ (1-x)^2\mu ^{2}}  \frac{d q_{\bot} ^{ 2}}{q_{\bot} ^{ 2}}   \int _{x}^{1-\frac{q_{\bot}}{\mu}} dz \Delta_a\left(\mu^{2} ,\frac{q_{\bot} ^{ 2}}{(1-z)^2}\right)   \sum _{b} P^{R}_{ab}\left(  \alpha_s\left(q_{\bot}^{ 2}\right), z\right)\widetilde{f}_{b}\left(\frac{x}{z},\frac{q_{\bot} ^{2}}{(1-z)^2}\right)  \;.
\end{eqnarray} 
The KMRW angular ordered evolution equation has the following form
\begin{eqnarray}\label{eq:KMR}
 && \widetilde{f}_{a}(x,\mu ^{2}) = \widetilde{f}_{a}(x,\mu _{0} ^{2})\Delta_a(\mu ^{2}, \mu_0 ^{2})  \nonumber \\  &&+ \int_{q_{0} ^{2}}^{ \mu^2\left(\frac{1-x}{x}\right)^2}  \frac{d q_{\bot} ^{ 2}}{q_{\bot} ^{ 2}}\left(\Delta_a(\mu ^{2}, q_{\bot} ^{ 2})\sum _{b} \int _{x}^{1-\frac{q_{\bot}}{q_{\bot}+ \mu}} dz P^{R}_{ab}\left(\alpha_s\left(q_{\bot}^{ 2}\right),z\right)\widetilde{f}_{b}\left(\frac{x}{z}, q_{\bot}^{ 2}\right) \right)
\end{eqnarray}
where the expression $
\widetilde{f}(x,\mu ^{2}, q_{\bot}^2)=\Delta_a(\mu ^{2}, q_{\bot} ^{ 2})\sum _{b} \int _{x}^{1-\frac{q_{\bot}}{q_{\bot}+ \mu}} dz P^{R}_{ab}\left(\alpha_s(q_{\bot} ^{ 2}),z\right)\widetilde{f}_{b}\left(\frac{x}{z}, q_{\bot}^{ 2}\right)
$ defines the TMD (called {\it{unintegrated PDF}} there).
KMRW is {\it{one-step}} evolution: it generates the second scale only in the last step of the evolution in contrast to PB method where both $k_{\bot}$ and $\mu^{\prime}$ (or equivalently $q_{\bot}$) are calculated at each branching.
Nevertheless, it is interesting to compare the formulas eq.~(\ref{PBangular_2term_3}) and eq.~(\ref{eq:KMR}).
First, we notice that  in KMRW and PB the integration limits differ. Moreover, PB and KMRW use  different scales in parton densities $\widetilde{f}_b$ and Sudakov form factors $\Delta_a$. Both formalisms use $q_{\bot}$ as the scale in $\alpha_s$. The TMD sets obtained according to KMRW angular ordering prescription are included in TMDlib and TMDplotter \cite{Hautmann:2014kza} under the name MRW-CT10nlo \cite{Bury:2017jxo}. Despite many differences, PB and KMRW are similar in the middle $k_{\bot}$ range compared to the scale $\mu$ which is illustrated in the middle and right panels of fig.~\ref{fig:KMRWvsPB} \footnote{PB TMDs were obtained with $q_0=1\textrm{GeV}$ and cut in $\alpha_s$ forbidding the renormalization scale to go below the initial evolution scale. The starting distribution is ct10nlo.}. They differ in the low $k_{\bot}$ region where for KMRW we see the  intrinsic $k_{\bot}$ constant parametrization whereas for PB the Gaussian intrinsic $k_{\bot}$  is smeared during the evolution process. PB and KMRW differ also in the large $k_{\bot}$ region: KMRW has a very large $k_{\bot}$ tail coming from their treatment of the Sudakov form factor for $k_{\bot}> \mu$.

\subsection{Collins-Soper-Sterman}
The PB Sudakov form factor  written in terms of virtual pieces of the splitting functions and with angular ordering  has the  form $\Delta_{a}(\mu ^{2},\mu_0^2)=\exp \left(- \int _{\mu _{0}^{2}}^{\mu ^{2}} \frac{d q_{\bot} ^{ 2}}{q_{\bot} ^{ 2}}\alpha_s(q_{\bot})\left(\int _{0}^{1-\frac{q_{\bot}}{\mu}} dz\left(k_a \frac{1}{1-z}\right)-d_a \right)\right)$. By comparing  the coefficients $k_q$ and $d_q$ \cite{Hautmann:2017fcj} in the PB Sudakov with $A_1$ (giving leading logarithmic (LL) contributions), $B_1$ and $A_2$ (giving together with $C_1$ next-to-leading logarithmic (NLL) contributions) coefficients of the CSS formalism for Drell-Yan (DY) cross section \cite{Collins:1984kg} we find that they are exactly the same. The fact that coherent branching algorithms should reproduce corretly LL and NLL behavior for DY and DIS was discussed in  \cite{Catani:1990rr}. 

It was realised \cite{Catani:2000vq}  that Sudakov form factor is process dependent what can be explained by renormalization group equation and resummation scheme dependence. We find a difference between $B_2$ CSS coefficient (giving together with $A_3$ and $C_2$ next-to-next-to leading (NNLL) logarithmic contribution) and the $2d_q^{1}$ coefficient in PB Sudakov  being $\pi\beta_0 16\left(\frac{\pi^2}{6}-1\right)$ where $\beta_0$ is the first coefficient of the expansion of the QCD $\beta$ function. 
\section{Conclusions}
PB method allows to obtain collinear and TMD PDFs by calculating the kinematics at each branching and to study different ordering definitions. We have shown that angular ordering definition leads to stable, $z_M$-independent  TMDs and good description of Z boson $p_{\bot}$ spectrum. 

We showed that PB method agrees with Marchesini's and Webber's  prescription. We discussed the differences and similarities of  PB and  KMRW approach. We illustrated that PB includes the same LL and NLL coefficients in the Sudakov form factor as CSS  formalism.  The differences between NNLL coefficients in the Sudakov form factors of these two methods come from the resummation scheme dependence.

\textbf{Acknowledgements}
The  presented results were obtained in collaboration with Francesco Hautmann,  Aron Mees van Kampen and  Lissa Keersmaekers. We thank Hannes Jung for many  discussions. 

\vspace{-0.5cm}
\section*{References}

\begin{thebibliography}{99}

\bibitem{Collins:1989gx}

  J.~C.~Collins, D.~E.~Soper and G.~F.~Sterman,
  Adv.Ser. Direct. High Energy Phys. {\bf 5} (1989) 1

\bibitem{Catani:1990xk}
  S.~Catani, M.~Ciafaloni and F.~Hautmann,
  Phys.\ Lett.\ B {\bf 242} (1990) 97

\bibitem{Catani:1990eg}
  S.~Catani, M.~Ciafaloni and F.~Hautmann,
  Nucl.\ Phys.\ B {\bf 366} (1991) 135




\bibitem{Collins:1984kg}
  J.~C.~Collins, D.~E.~Soper and G.~F.~Sterman,
  Nucl.\ Phys.\ B {\bf 250} (1985) 199


\bibitem{Angeles-Martinez:2015sea}
  R.~Angeles-Martinez {\it et al.},
  Acta Phys.\ Polon.\ B {\bf 46} (2015) no.12,  2501
  
  
  
  

\bibitem{Hautmann:2017xtx}
  F.~Hautmann, H.~Jung, A.~Lelek, V.~Radescu and R.~Zlebcik,
  Phys. Lett. B {\bf 772} (2017) 446
  
  
\bibitem{Hautmann:2017fcj}
  F.~Hautmann, H.~Jung, A.~Lelek, V.~Radescu and R.~Zlebcik,
  JHEP {\bf 1801} (2018) 070

\bibitem{Martinez:2018jxt}
  A.~Bermudez Martinez, P.~Connor, H.~Jung, A.~Lelek, R.~Zlebcik, F.~Hautmann and V.~Radescu,
  Phys.\ Rev.\ D {\bf 99} (2019) no.7,  074008

\bibitem{Lelek:2018vtr}
  A.~Lelek,
  doi:10.3204/PUBDB-2018-02949
  
  
\bibitem{Hautmann:2019biw}
  F.~Hautmann, L.~Keersmaekers, A.~Lelek and A.~M.~Van Kampen,
  arXiv:1908.08524
  
  

  
  
\bibitem{Marchesini:1987cf}
  G.~Marchesini and B.~R.~Webber,
  Nucl.\ Phys.\ B {\bf 310} (1988) 461
  
    
\bibitem{Kimber:1999xc}
  M.~A.~Kimber, A.~D.~Martin and M.~G.~Ryskin,
  Eur.\ Phys.\ J.\ C {\bf 12} (2000) 655
  
  

 
 
\bibitem{Martin:2009ii}
  A.~D.~Martin, M.~G.~Ryskin and G.~Watt,
  Eur.\ Phys.\ J.\ C {\bf 66} (2010) 163
 
  

  
  
\bibitem{Gribov:1972ri}
  V.~N.~Gribov and L.~N.~Lipatov,
  Sov.\ J.\ Nucl.\ Phys.\  {\bf 15} (1972) 438
  
\bibitem{Lipatov:1974qm}
  L.~N.~Lipatov,
  Sov.\ J.\ Nucl.\ Phys.\  {\bf 20} (1975) 94
  
\bibitem{Altarelli:1977zs}
  G.~Altarelli and G.~Parisi,
  Nucl.\ Phys.\ B {\bf 126} (1977) 298
  
  
\bibitem{Dokshitzer:1977sg}
  Y.~L.~Dokshitzer,
  Sov.\ Phys.\ JETP {\bf 46} (1977) 641
  
 
  
  
\bibitem{Alekhin:2014irh}
  S.~Alekhin {\it et al.},
  Eur.\ Phys.\ J.\ C {\bf 75} (2015) no.7,  304



  
\bibitem{Aad:2015auj}
  G.~Aad {\it et al.} [ATLAS Collaboration],
  Eur.\ Phys.\ J.\ C {\bf 76} (2016) no.5,  291
  
  
\bibitem{Martinez:2019mwt} 
  A.~Bermudez Martinez {\it et al.},
  Phys.\ Rev.\ D {\bf 100}, 074027 (2019)
  
  
\bibitem{Hautmann:2014kza}
  F.~Hautmann, H.~Jung, M.~Krämer, P.~J.~Mulders, E.~R.~Nocera, T.~C.~Rogers and A.~Signori,
  Eur.\ Phys.\ J.\ C {\bf 74} (2014) 3220
  
\bibitem{Bury:2017jxo}
  M.~Bury, A.~van Hameren, H.~Jung, K.~Kutak, S.~Sapeta and M.~Serino,
  Eur.\ Phys.\ J.\ C {\bf 78} (2018) no.2,  137
  
  
\bibitem{Catani:1990rr}
 S.~Catani, B.~R.~Webber and G.~Marchesini,
 Nucl.\ Phys.\ B {\bf 349} (1991) 635
  
  
  
\bibitem{Catani:2000vq}
  S.~Catani, D.~de Florian and M.~Grazzini,
  Nucl.\ Phys.\ B {\bf 596} (2001) 299
  
\end{thebibliography}
\end{document}